\documentclass[preprint,epsfig,aps,showpacs]{revtex4}
\usepackage{amsmath}
\usepackage{graphicx}

\newcommand{\f}{\begin{equation}}
\newcommand{\ff}{\end{equation}}
\newcommand{\fa}{\begin{eqnarray}}
\newcommand{\ffa}{\end{eqnarray}}

\begin{document}
\title{The power-law expansion universe and dark energy evolution}
\author{Yi-Huan Wei${ }^{1,2}$}
\affiliation{%
${ }^1$ Department of Physics, Bohai University, Jinzhou 121000,
 Liaoning, China}
\affiliation{%
$ { }^2$ Institute of Theoretical Physics, Chinese Academy of
Science, Beijing 100080, China}
\begin{abstract}
~~In order to depict the transition from deceleration to
acceleration expansion of the universe we use a power-law
expansion scale factor, $a\sim t^{n_0+bt^m}$, with $n_0$, $b$ and
$m$ three parameters determined by $H_0$, $q_0$ and $z_T$. For the
spatially flat, isotropic and homogeneous universe, such a scale
factor leads to the results that the dark energy density is slowly
changing currently, and predicts the equation of state $w_X$
changes from $w_X>-1$ to $w_X<-1$.

\end{abstract}

\maketitle

\section {Introduction}

The type Ia supernova (SNe) observations \cite{Riess} and the
measurements on the cosmic microwave background (CMB) \cite{Ber}
indicate the universe is acceleratedly expanding and is spatially
flat. The faintness of high-redshift SNe has been interpreted as
evidence that the expansion of the universe is accelerating, which
suggests the universe transition from deceleration to acceleration
has happened. The transition redshift of the universe from
deceleration to acceleration provides us an important information
of dark energy. Providing that the apparent faintness of SNe Ia
provides the direct evidence for the accelerating expansion of the
universe, then it happened at $z_T\approx 0.5$ \cite{Turn}. It
should have happened at $z>0.3$, with the best fit at about
$z_T\approx 0.45$ \cite{Daly}, or $z_T\approx 0.40$ \cite{Dal}.
Using a simple expansion model, $q\simeq q_0+z(dq/dz)_{z=0}$, and
the statistical method, the transition redshift is determined to
be $z_T\simeq0.46$ \cite{Rie}. From modified Friedmann equation,
the constraint from the SNe Ia observation gives
$z_T=0.52\sim0.73$ \cite{Zhu}. From a joint analysis of SNe+CMB,
one can get $z_T\approx0.39$, but it can be $z_T\approx0.57$ for
$\Omega_{0m}\approx0.27$, $h\approx0.71$ and $w_\Lambda\approx-1$
\cite{Alam}. Anyway, the results from astronomical observations
show our universe could have undergone an expansion transition
from deceleration to acceleration
\cite{Turn,Daly,Dal,Rie,Zhu,Alam,Ries}.

The problem of the dynamical origin of the dark energy has
attracted a great many attentions
\cite{CRS,K,T,C,Caldwell,Wei,Wei1,Guo,P,Odintsov}. Current
supernova, CMB, and LSS data already rule out dark energy models
with dark energy densities varying fast with time \cite{Yun}. From
the viewpoint of geometry the scale factor of the universe is a
fundamental quantity. It is possible to catch a glimpse of the
nature of dark energy. Rather than attempting to obtain a whole
description for the universe and the dark energy, the approach
here aims to construct a scale factor by using the current
observations: the Hubble parameter $H_0$, the deceleration
parameter $q_0$ and the transition redshift. It is expected that
such a scale factor provides an approximate descriptions for the
universe and the dark energy during a history of evolution
including at least the transition of the universe.

The determination of parameters in a parametrized quantity depends
on the observations. An appropriate scale factor should satisfy at
least the two requirements from the observations: the past
transition of the universe from acceleration to deceleration
expansion phase and the currently slow variation of the dark
energy density. Recently, motivated by describing the possible
super-accelerated transition \cite{Wei}, we discuss the scalar
factor of the form $a\sim t^{n_0+bt}$. In this paper, we will
extend this form to $a\sim t^{n_0+bt^m}$. This needs the three
observation quantities to determine the three parameters, which
will be chosen to be $H_0$, $q_0$ and $z_T$. In Sec. II, we
determine the three parameters in the scale factor by using
$H_0\approx0.7\times 10^{-10}yr^{-1}$, $q_0=-0.5$ and $z_T=0.5$,
as well as $t_0\approx1.4\times 10^{10}yr$. In Sec. III, we show
the results of dark energy produced from the above scale factor
with three parameters, which illustrates that the dark energy
density is slowly changing currently and the equation of state
$w_X$ can change from $w_X>-1$ to $w_X<-1$. The latter result can
give a strict constraint on the dark energy model.

\section {Approach on the power-law expansion universe}

For each epoch, the radiation, matter and dark energy epoch, the
universe may be characterized by a power-law scale factor with a
constant exponent depending on the equation of state of the
dominant component in that epoch. Actually, the exponent change
with time since the fractions of the different energy components
in the universe have been varying. We begin by writing down the
scale factor
\begin{eqnarray}
a=a_c(\frac{t}{T})^n, \label{eq1}
\end{eqnarray}
with $n$ a function of time, where $a_c$ and $T$ are two
constants. Ftom (\ref{eq1}), the Hubble parameter and its
derivative are
\begin{eqnarray}
H=\dot{n}\ln t+\frac{n}{t}, \label{eq2}
\end{eqnarray}
\begin{eqnarray}
\dot{H}=\ddot{n}\ln t+2\frac{\dot{n}}{t}-\frac{n}{t^2},
\label{eq3}
\end{eqnarray}
where a dot denotes the derivative with respect to time.

Without observations, there is no priori way of determining the
scale factor. We anticipate, from our knowledge of the evolution
of the universe such as the existence of the transition, that the
exponent should increase with time at least from the
matter-dominated epoch to today. Assuming the scale factor
(\ref{eq1}), the simplest form of the exponent to satisfy the
above requirement should be the linear approximation $n=n_0+bt$,
as in \cite{Wei}, which can always be a good approximation to the
regular exponent $n$ in an enough short time. However, we require
a scale factor can well describe the universe in a long interval
of time, here. For this, we will attempt to test the following
form of the scale factor
\begin{eqnarray}
n=n_0+bt^m, \label{eq4}
\end{eqnarray}
where $n_0$, $m$ and $b$ are three positive constants.

Putting Eq.(4) and $\dot{n}=bmt^{m-1}$ in Eqs. (\ref{eq2}) and
(\ref{eq3}) yields
\begin{eqnarray}
H=b(m\ln{t}+1)t^{m-1}+n_0t^{-1}, \label{eq5}
\end{eqnarray}
\begin{eqnarray}
\dot{H}=b[m(m-1)\ln{t}+2m-1]t^{m-2}-n_0t^{-2}. \label{eq6}
\end{eqnarray}
Eqs. (\ref{eq5}) and (\ref{eq6}) shows the dependence of $H$ and
$\dot{H}$ on $m$. For $m<1$, Eq. (\ref{eq6}) leads to $\dot{H}$ is
always negative and (\ref{eq5}) indicates the universe gets
gradually flat for late time. For $m\geq1$, Eqs. (\ref{eq5}) and
(\ref{eq6}) show obviously the late-time phantom property of the
dark energy \cite{Wei}.

From $\ddot{a}/a=\dot{H}+H^2$ and $H=\dot{a}/a$, one has the
deceleration parameter
\begin{eqnarray}
q=-\frac{a\ddot{a}}{\dot{a}^2}=-(1+\frac{\dot{H}}{H^2}).
\label{eq7}
\end{eqnarray}
Defining $x=\ln{t}$, $A=2n_0+2m-1$ and $B=m(2n_0+m-1)$, putting
Eqs. (\ref{eq2}) and (\ref{eq3}) in (\ref{eq7}) then we obtain
\begin{eqnarray}
q= -H^{-2}[b^2(1+mx)^2t^{2m-2}+b(A+ Bx)t^{m-2}+(n_0^2-n_0)t^{-2}].
\label{eq8}
\end{eqnarray}

Using $H_0$, $q_0$, and $t_0$, then from Eqs. (\ref{eq5}) and
(\ref{eq8}) we obtain the following two equations
\begin{eqnarray}
H_0=b(mx_0+1)t_0^{m-1}+n_0t_0^{-1}, \label{eq9}
\end{eqnarray}
\begin{eqnarray}
-q_0H_0^2= b^2(1+mx_0)^2t_0^{2m-2}+b(A+
Bx_0)t_0^{m-2}+(n_0^2-n_0)t_0^{-2}, \label{eq10}
\end{eqnarray}
with $x_0=\ln{t_0}$. Solving Eqs. (\ref{eq9}) and (\ref{eq10})
gives rise to
\begin{eqnarray}
n_0=H_0t_0-b(mx_0+1)t_0^m, \quad
b=\frac{1-(q_0+1)H_0t_0}{m(mx_0+2)}H_0t_0^{1-m}. \label{eq11}
\end{eqnarray}

\begin {figure}[!hbt]
\begin{center}
\includegraphics{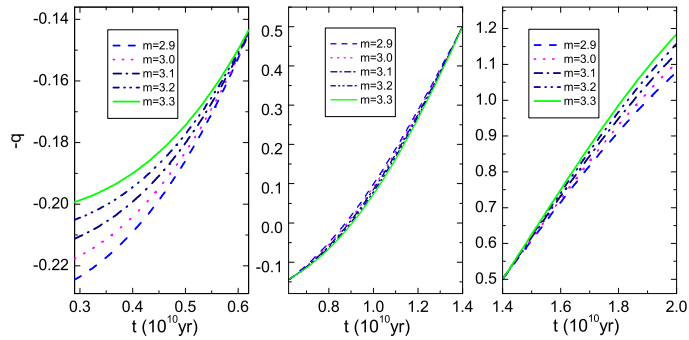}
\caption{The figure shows the variations of $q$ for $m=2.90, 3.00,
3.10, 3.20, 3.30 $, respectively.} \label{Fig.1}
\end{center}
\end{figure}

According to \cite{Dal}, there can be the current deceleration
parameter $q_0>-0.5$. From \cite{Rie,Alam}, there can be
$q_0<-0.5$. Let us adopt a modest value $q_0=-0.5$, here. The
current Hubble parameter and the cosmological age are taken as
$H_0\approx0.7\times 10^{-10}yr^{-1}$ and $t_0\approx1.4\times
10^{10}yr$ \cite{Freedman}, here. The three parameters
($H_0\approx0.7\times 10^{-10}yr^{-1},q_0=-0.5,
 z_T=0.5$), as well as,
$t_0\approx1.4\times 10^{10}yr$ will be used to determine $n_0$,
$b$ and $m$. For some special $m$, the parameters $b$ and $n_0$
are exhibited in Table \ref{tab1}. Notice that the deceleration
parameter $q$ is given by (\ref{eq7}). For the five cases
$m=2.9\sim3.3$, the variations of the deceleration parameter $q$
with time are shown in Fig. \ref{Fig.1}. From Fig. \ref{Fig.1},
one can see the expected converge of the curves characterized by
$m$ to $t=t_0$ and the unexpected converge to
$t\approx0.6\times10^{10}yr$. The second convergence of the curves
with different $m$ is an intriguing feature, it means the validity
of the power-law scale factor may be extended to the range,
$t<0.6\times10^{10}yr$.
\begin{table}[!hbt]
\begin{center}
\caption{The parameters $b$, $n_0$ are given for $q_0=-0.50$ and
some special values of $m$.} \vskip 0.3cm
\begin{tabular}{c|c|c|c|c|c}
\hline\hline
 \quad $m$ \quad & $2.90$ & $3.00$ & $3.10$ & $3.20$ &
$3.30$  \\
\hline \quad $b$ \quad  & $9.32\times 10^{-33}$ & $8.42\times
10^{-34}$ & $7.63\times
 10^{-35}$ & $6.93\times 10^{-36}$ & $6.31\times 10^{-37}$ \\
 \hline
 \quad $n_0$ \quad & $0.810$ & $0.816$ & $0.821$ & $0.826$ & $0.830$ \\
 \hline\hline
\end{tabular}
\label{tab1}
\end{center}
\end{table}

The expression, $1+z=a_0/a(t)$, holds valid for any spatially
flat, isotropic and homogeneous universe. Table 2 exhibits the
predicted transition time $t_T=8.75\sim9.06Gyr$ for $m=2.9\sim
3.3$. From $z_T=t_0^{n_1}/t_T^{n_2}-1$ with $n_1=n_0+bt_0^m$ and
$n_2=n_0+bt_T^m$, the corresponding transition redshift is
$z_T=0.528\sim0.485$. For $z_T=0.5$, $m$ is determined as
$m\approx3.16$, through (\ref{eq11}) which gives $n_0\simeq0.8239$
and $b\simeq1.809\times10^{-35}yr^{-m}$.
\begin{table}[!hbt]
\begin{center}
\caption{The predicted transition time $t_T$ and the corresponding
redshift $z_T$.} \label{tab2} \vskip 0.3cm
\begin{tabular}{c|c|c|c|c|c}
\hline\hline
 $m$ &$2.90$ & $3.00$ & $3.10$ & $3.20$ &
$3.30$  \\
\hline
 $t_T(10^{10}yr)$  & $0.875$ & $0.883$ & $0.891$ & $0.8984$ & $0.906$  \\
 \hline
 $z_T$ & $0.528$ & $0.517$ & $0.506$ & $0.496$ &
 $0.485$ \\
 \hline\hline
\end{tabular}
\end{center}
\end{table}
\begin{table}
\begin{center}
\caption{This table shows $\triangle n=bt^m$ with
$b\simeq1.809\times10^{-35}yr^{-m}$ and $m=3.16$.} \label{tab3}
\vskip 0.3cm
\begin{tabular}{c|c|c|c|c}
\hline\hline
$t(10^{10}yr)$  & $0.6$ & $0.9$ & $1.4$ & $2.0$   \\
\hline
 $\triangle n$ & $0.0001$ & $0.0005$ & $0.0021$ & $0.0064$  \\
 \hline\hline
\end{tabular}
\end{center}
\end{table}

TABLE III shows the variation of $n$ with time is very slow, which
changes only about $0.0064$ or $n\simeq 0.8239\sim0.8303$ in
$t=6Gyr\sim20Gyr$. This illustrates that the scale factor
discussed currently can be applicable from the late
matter-dominated epoch. In this section, we have worked out the
three parameters $n_0$, $b$ and $m$ by using the three observation
quantities $H_0$, $q_0$ and $z_T$. In the next section, we will
extract some some information of the dark energy from the scale
factor with the above three parameters.

\section {Resulting dark energy evolution from scale factor with three parameters}

Though the scale factor (\ref{eq1}) with the exponent (\ref{eq4})
is somewhat an ad hoc choice, it has taken into account the actual
observations such as the transition of universe from acceleration
to deceleration expansion. It is expected that the power-law
expansion universe proposed here can yield the consistent result
with the observations for the dark energy.

For the spatially flat, isotropic and homogeneous universe, the
Einstein equations reads
\begin{eqnarray}
(\frac{\dot{a}}{a})^2=\frac{3}{8\pi G}(\rho_X+\rho_M),
\label{eq-F1}
\end{eqnarray}
with $\rho_X$ and $\rho_M$ the energy densities of dark energy and
matter. For the matter of the pressureless fluid, the energy
density evolves in terms of
\begin{eqnarray}
\rho_{M}=\rho_{M0}(1+z)^3=\Omega_{M0}\rho_0\frac{t_0^{3(n_0+bt_0^m)}}{t^{3(n_0+bt^m)}},
\label{eqm}
\end{eqnarray}
where $\rho_{0}$ is the total energy density of the current
universe, $\rho_{M0}$ and $\Omega_{M0}$ denote the today's energy
density and fraction of the matter. From Eqs. (\ref{eq-F1}) and
(\ref{eqm}), it follows that the energy density of dark energy
\begin{eqnarray}
\rho_X=\rho-\rho_M=[\frac{[n_0+b(m\ln{t}+1)t^{m}]^2}{H_0^2t^2}-
\Omega_{M0}\frac{t_0^{3(n_0+bt_0^m)}}{t^{3(n_0+bt^m)}}]\rho_0.
\label{eqmp}
\end{eqnarray}

\begin {figure}[!hbt]
\begin{center}
\includegraphics{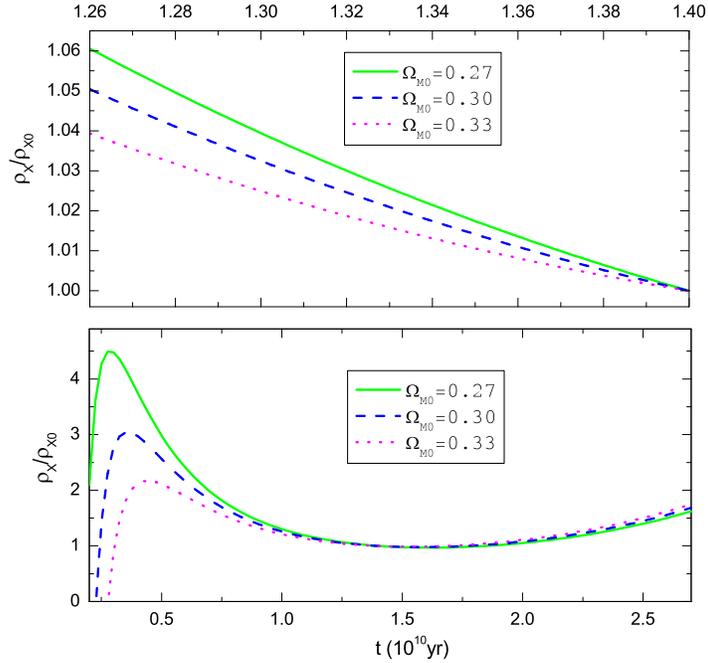}
\caption{ Dark energy evolution under the three cases,
$\Omega_{M0}=0.27,0.30,0.33$, respectively.} \label{Fig.2}
\end{center}
\end{figure}

Fig. 2 shows how $\rho_X$ varies with time. From Fig. 2, one can
see that from $t=12.6Gyr$ ($z\simeq 0.1$) to $t=14Gyr$ the dark
energy density changes only about $6\%\sim4\%$ for
$\Omega_M=0.27\sim0.33$, this illustrates that the dark energy is
slowly changing currently \cite{Yun,Jas}. The conserved equation
for dark energy reads
\begin{eqnarray}
\dot{\rho}_X+3H(\rho_X+p_X)=0. \label{eq15}
\end{eqnarray}
Noting that the equation of state $p_X=w_X\rho_X$, Eq.
(\ref{eq15}) yields the approximate calculation formula of the
equation of state, $\bar{w}_X=-1-\frac{\triangle
\rho_X}{3\bar{H}\bar{\rho}_X\triangle t}$, where $\dot{\rho}_X$,
$H$ and $\rho_X$ have been replaced by the mean variation rate
$\frac{\triangle \rho_X}{\triangle t}$, the mean values $\bar{H}$
and $\bar{\rho}_X$ for $\triangle t\ll 1Gyr$, respectively. From
the $\bar{w}_X$ formula, then there is
\begin{eqnarray}
\bar{w}_{X0}=-1-\frac{\triangle \rho_X}{3H_0\rho_{X0}\triangle t},
\label{eq16}
\end{eqnarray}
where $\bar{w}_{X0}$ denotes the average of $w_X$ in $\triangle
t=t_0-t$ with $t_0$ the age of the universe. Taking $t=13.9Gyr$,
i.e., $\triangle t=0.1Gyr$, then
$\frac{\triangle\rho_X}{\rho_{X0}}\simeq -0.0031,-0.0025,-0.0018$,
and (\ref{eq16}) yields $w_{X0}=\bar{w}_{X0}\simeq -0.85, -0.88,
-0.92$ for $\Omega_{M0}=0.27,0.30,0.33$, respectively. Clearly,
the above values of $w_{X0}$ can agree with the most results of
the dark energy equation of state.

In addition, Fig. 2 illustrates $\rho_X$ has the minimum, which
implies that $w_X$ can change from $w_X>-1$ to $w_X<-1$ at a
certain future time, which is quite similar to the result
\cite{Yun,Huterer} that $w_X$ can be across $-1$ at $z<0.2$.
Looking somewhat different, these two results don't mean any
disagreement since the most observations such as the equation of
state of dark energy are still given only in a quite uncertainty
range. Instead, it can provide an important information of the
dark energy. If getting confirmed from the further observations it
will provide a strong constrain on the dark energy models.

To sum up, the predicted results of the dark energy from the
parametrized scale factor with three parameters show no
contradiction with the known those from observations. Thus, it is
believed that it can give a good approximation descriptions for
the universe and dark energy during a long evolution course, from
$6Gyr$ to a future time, saying $20Gyr$. We hope that the work in
this paper can be the beginning of a meaningful approach on the
universe evolution and the behavior of dark energy.

\vskip 0.9cm

{\bf Acknowledgement:}  This work is supported by Liaoning
Province Educational Committee Research Project (20040026) and
National Nature Science Foundation of China; it was supported ITP
Post-Doctor Project (22B580), Chinese Academy of Science of China.

\end{document}